\documentclass[pra,aps,showpacs,superscriptaddress,amsmath,amssymb,floatfix,twocolumn]{revtex4}
\usepackage{graphicx}
\usepackage{color}

\begin{document}

\title{Heisenberg Uncertainty Principle as Probe of Entanglement Entropy: Application to Superradiant Quantum Phase Transitions\footnote{pierre.nataf@epfl.ch  \\$^{\dag}$ karyn.lehur@cpht.polytechnique.fr}}
\date{\today} 

\author{Pierre Nataf,$^{1,2,*}$ Mehmet Dogan$^{2}$, and Karyn Le Hur $^{4,3,\dag}$ \\ 
\emph{$^{1}$Institute of Theoretical Physics, Ecole Polytechnique F\'ed\'erale de Lausanne (EPFL), CH-1015 Lausanne, Switzerland\\
$^{2}$Laboratoire Mat\'eriaux et Ph\'enom\`enes Quantiques,
Universit\'e Paris Diderot-Paris 7 et CNRS, \\ B\^atiment Condorcet, 10 rue
Alice Domon et L\'eonie Duquet, 75205 Paris Cedex 13, France\\
$^{3}$ Department of Physics, Yale University, New Haven, Connecticut 06520, USA\\
$^{4}$Centre de Physique Th\'{e}orique, Ecole Polytechnique, CNRS, 91128 Palaiseau Cedex, France }}

\begin{abstract}
Quantum phase transitions are often embodied by the critical behavior of purely quantum quantities such as entanglement or quantum fluctuations.
In critical regions, we underline a general scaling relation between the entanglement entropy and  one of the  most fundamental and simplest measure of the quantum fluctuations, the Heisenberg uncertainty principle. Then, we show that the latter represents a sensitive probe of superradiant quantum phase transitions in standard models of photons such as the Dicke Hamiltonian, which embodies an ensemble of two-level systems interacting with one quadrature of a single and uniform bosonic field. We derive exact results in the thermodynamic limit and for a finite number $N$ of two-level systems:  as a reminiscence of the entanglement properties between light and the two-level systems,  the product $\Delta x \Delta p$ diverges at the quantum critical point as $N^{1/6}$. We generalize our results to the double quadrature Dicke model where the two quadratures of the bosonic field are now coupled to two independent sets of two-level systems. Our findings, which show that the entanglement properties between light and matter can be accessed through the Heisenberg uncertainty principle, can be tested using Bose-Einstein condensates in optical cavities and circuit quantum electrodynamics.
 \end{abstract}

     \maketitle

\section{Introduction}
Quantum phase transitions (QPTs) occur at zero temperature and are triggered by quantum fluctuations \cite{sachdev}. 
They manifest themselves in a sudden change of quantum properties of the collective ground states. 
The appearance of a finite order parameter or collective gapless excitation is often used as a signature of such a critical phenomenon. Recently, tools of quantum informations were also used in this perspective. For instance, the Von Neumann entanglement entropy \cite{review} exhibits a critical behavior in many collective models undergoing a QPT \cite{collective,lambert,spinboson}. 
Other examples of such quantities can appropriately locate the Quantum critical points (QCP) of some given  systems.
 Among them, one can quote  the  fidelity (or overlap)  \cite{zanardi} which directly quantifies the suddenness of the change of the ground state wave-function around the QCP, the global geometric entanglement \cite{geometric} which evaluates its distance, in the Hilbert space, to the closest separable state, the logarithmic negativity \cite{negativity}, which measures the entanglement between non-complementary parts of the system,  and more recently, the bipartite fluctuations which have been shown \cite{bipartite} to  be directly related to the entanglement properties between subsystems of some one- or higher- dimensional fermionic models.\\

This article is aimed to demonstrate that the critical behavior of the Heisenberg principle (HP) can be used as a characterization of QPTs as well as entanglement properties, in particular in systems of photons. 

Since QPTs are driven by quantum fluctuations of the fields that define the physical structure of the system, it is rather natural to focus on the simplest  measure of those fluctuations, the product $\Delta x \Delta p$. Secondly, and importantly, fluctuations appear in the calculation of the entanglement of many important physical models \cite{bombelli}. 

In the case of light-matter systems, one can derive the Von Neumann entropy of  the reduced density matrix $\rho_r$ of the photon mode 
$S\equiv-\hbox{Tr}\{\rho_r \log_2 (\rho_r)\}$. The latter is obtained at zero temperature  by tracing out the matter degrees of freedom in the density matrix $\rho=|G\rangle\langle G|$ (where $|G\rangle$ is the collective ground state), so that $\rho_r=\hbox{Tr}_{\hbox{matter}}\{|G\rangle\langle G|\}$. Then,  providing that the model can be described by a quadratic Hamiltonian of interacting bosonic fields \cite{barthel}, the entanglement entropy S will be directly related to the HP through:
\begin{align}
\label{relation}
S&=-\hbox{Tr}\{\rho_r \log_2 (\rho_r)\}\nonumber\\
&=\left(\Delta x\Delta p\,+\,\frac{1}{2}  \right) \log_2\left \{\Delta x\Delta p\,+\,\frac{1}{2} \right\} \nonumber\\
&- \left(\Delta x\Delta p\,-\,\frac{1}{2} \right)  \log_2\left \{\Delta x\Delta p\,-\,\frac{1}{2} \right\},
\end{align}  
with $\Delta x\Delta p=(1/2)\sqrt{\langle G |(a+a^{\dag})^2| G\rangle\langle G|-(a-a^{\dag})^2| G\rangle}$ ($\hbar=1$).
Since we consider that this equality constitutes an important relation between entanglement properties of the system and fluctuations, for completeness, we shall derive it below, in Sec. \ref{review}. From this equation, one even obtains the very simple relation: 
\begin{align}
\label{relentang}
S \sim \log_2\{ \Delta x\Delta p \}\,\,\,\,\,\,\,\,\,\,\hbox{if}\,\,\,\,\,\,\,\,\Delta x\Delta p \gg 1.
\end{align} 
 Thus, a divergence of such a quantity would directly imply a logarithmic divergence of the entanglement entropy.
 The $\alpha$-R\'{e}nyi entropies, which can individuate a QCP \cite{romera}, are also  very sensitive to the divergence of the HP.
Finally, the individuation of a QPT through the HP seems to be an important concept to push forward owing to the fact that such quantity is directly accessible experimentally, especially in the context of quantum optics. Incidentally, it is remarkable to notice all the efforts made so far to develop methods for determining the entanglement in many-body systems from the measure of physical observables \cite{efforts}.\\
 
 In the following, we will investigate two simple yet relevant models exhibiting QPTs.
First, the HP will be investigated in the case of the celebrated Dicke model \cite{Dicke}, where a bosonic field interacts through one of its quadratures with a chain of $N$ two-level systems. Then, the double quadrature Dicke model \cite{dblDicke}, where the two quadratures of the bosonic field are coupled to two independent chains of atoms, will be analyzed thoroughly. This recent model provides a very natural framework for the investigation of the critical behavior of the product  $\Delta x \Delta p$ at the QCPs of light models, because the fluctuations of the quadratures $x$ and $p$, which both  interact with two different sets of  atoms, play an equivalent role and can both undergo a critical enhancement at the QPT. Finally, possible applications of those models using Bose-Einstein condensates in optical cavities \cite{kimble,Zurich,ritsch} or circuit QED \cite{brune,buisson,wallraff,esteve} will be briefly discussed. \\

The paper is organized as follows. In Sec. \ref{review}, we will review the method of calculating entanglement entropies in quadratic Hamiltonians of interacting bosonic fields in order to demonstrate
Eqs. (\ref{relation}) and (\ref{relentang}). Then, in
Sec. \ref{dicke}, we will explore the criticality of the photonic fluctuations in the Dicke model, both in the thermodynamical limit case and for a finite number $N$ of two-level systems.
In Sec. \ref{dbldicke}, the fluctuations and the entanglement in the double quadrature Dicke model will be investigated, with a special emphasis on the point of double symmetry breaking where those quantities behave in a peculiar way. Finally, we conclude in Sec. \ref{conclusion}.

 \section{Relation between the Entanglement Entropy and the fluctuations in quadratic models}
 \label{review}
 
 We review here the derivation of the entanglement entropy and the $\alpha$-R\'{e}nyi entropies for the ground state of a quadratic Hamiltonian $\mathcal{H}(a,b_2,b_3,...,b_{n})$ expressed in terms of  $n$ different bosonic fields $a$, $b_2,b_3,...,b_{n}$,   by using the method introduced in the articles \cite{barthel}.
 Note that the following method can also be used for fermionic modes \cite{barthel}.
 First of all, it is well-known that the ground state of any quadratic bosonic Hamiltonian can be written as:
 \begin{align}
 |G\rangle=\mathcal{N} e^{(\Phi^{\dag})^T \mathcal{A}\Phi^{\dag}} |0\rangle \otimes |0\rangle..|0\rangle,
 \end{align}
 where $(\Phi^{\dag})^T=(a^{\dag},b_2^{\dag},b_3^{\dag},...,b_{n}^{\dag}$),  $\mathcal{A}$ is a symmetric (real or complex) $n\times n$ matrix , $\mathcal{N}$ is a normalization constant (for instance real), and $|0\rangle \otimes |0\rangle..|0\rangle$ is the vacuum of the fields
 $a$, $b_2,b_3,...,b_{n}$ : $a|0\rangle \otimes |0\rangle..|0\rangle=0$ and $b_i|0\rangle \otimes |0\rangle..|0\rangle=0$ for $i=2...n$.
 We now derive the reduced density matrix of the photon mode $a$ by tracing out the matter degrees of freedom $b_2,b_3,...,b_{n}$ in the pure ground state density matrix $\rho=|G\rangle\langle G |$: 
 \begin{widetext}
 \begin{align}
 \rho_r&=\hbox{Tr}_{\hbox{matter}}\{\rho\} = \hbox{Tr}_{b_2,..b_{n}}\{|G\rangle\langle G|\} \\\nonumber &= \mathcal{N}^2\iiiint \frac{d\nu_2 d\nu_3...d\nu_{n} }{\pi^{n-1}} \langle \nu_2 | \otimes \langle \nu_3 |...\otimes \langle \nu_{n} |  e^{(\Phi^{\dag})^T \mathcal{A}\Phi^{\dag}} |0\rangle \otimes |0\rangle..|0\rangle 
\langle 0 | \otimes \langle 0 |..\langle 0 | e^{(\Phi)^T \mathcal{A}^*\Phi } | \nu_2 \rangle \otimes | \nu_{3} \rangle ...\otimes | \nu_{n} \rangle \nonumber \\
&=\mathcal{N}^2 e^{ \mathcal{A}_{11} (a^{\dag})^2}\Bigg{(}\int\frac{d\nu_{i_0}}{\pi}e^{(2\sum_{i=2}^n\mathcal{A}_{1i} \nu_i^* )a^{\dag}}|0\rangle 
\langle 0|e^{(2\sum_{i=2}^n \mathcal{A}_{1i}^* \nu_i )a}  \iiiint \frac{\Pi_{i\neq i_0}d\nu_i}{\pi^{n-2}} e^{\sum_{i,j=2}^n  (\mathcal{A}_{ij}\nu_i^*\nu_j^*+ h.c)-\delta_{i,j}|\nu_i|^2  } \Bigg{)} e^{ \mathcal{A}_{11}^* a^2}\nonumber,
 \end{align}
   \end{widetext}
 where for $i=2...\,n$, the states $|\nu_i\rangle$ are coherent states\cite{gazeau} for the $i^{th}$ field: $|\nu_i\rangle = e^{-\frac{|\nu_i^2|}{2}}e^{\nu_i b_i^{\dag}} |0\rangle$ (satisfying $b_i |\nu_i\rangle = \nu_i |\nu_i\rangle$). The index $i_0$ is such that $\mathcal{A}_{1i_0}\neq 0$ (if such an index does not exist, the problem becomes trivial).
Then, we introduce the new variables $\nu'_i=\nu_i$ ($i\neq i_0$) and $\nu'_{i_0}=2\sum_{i=2}^n \mathcal{A}_{1i}^* \nu_i$.
 Now since the integral of a gaussian is a gaussian,  and using that $ \int\frac{d\nu_{i_0} }{\pi}e^{|\nu_{i_0}|^2(1-\frac{1}{|X|})} |\nu_{i_0}\rangle_a\langle\nu_{i_0}|_a=|X|^{a^{\dag}a+1}$ and $ \int\frac{d\nu_{i_0} }{\pi}e^{Y\nu_{i_0}^2} |\nu_{i_0}\rangle_a\langle\nu_{i_0}|_a=e^{Y a^2}$ for any number $X$ and $Y$, one can realize that $\rho_r$ is proportionnal to the product
 of several exponentials of $a^2$,  $(a^{\dag})^2$ and $a^{\dag} a$. One concludes \cite{bch2} that $\rho_r$ is the exponential of a quadratic Hermitian form in the field $a$ :
 \begin{align}
\rho_r=\exp\{-\kappa_0-\kappa_1 a^{\dag}a-\kappa_2 a^2-\kappa_2^* (a^{\dag})^2\},
 \end{align}
 where $\kappa_0$ and $\kappa_1$ are real, and $\kappa_2$ {\it a priori} complex.
 One  diagonalizes $-\{\kappa_0+\kappa_1 a^{\dag}a+\kappa_2 a^2+\kappa_2^* (a^{\dag})^2\}$ by a Bogoliubov transformation
 to write :
 
 \begin{align}
\rho_r=\exp\{-E_0-\Delta P^{\dag}P\},
 \end{align}
 where $P=u a+ v a ^{\dag}$, with $u$ and $v$ such that $|u|^2-|v|^2=1$, (and $u$ chosen to be real), where $\Delta$ is the pseudo-energy (not to be confused with the fluctuations $\Delta x$ or $\Delta p$), and where $E_0$ is a constant.
 Then, since  $\hbox{Tr}_{a} (\rho_r A)=\hbox{Tr}_{a,b_2,..b_{n}} (A\rho)=\langle G |A|G \rangle $
 for any photonic operator $A$, it is easy to determine $E_0$ , $\Delta$,  $u$, and $v$.
 Inverting the Bogoliubov transformation leads to $a=u^*P-vP^{\dag}$ and allows us to write:
\begin{align}
\left(\frac{1+e^{-\Delta}}{1-e^{-\Delta}}\right)^2&=(\langle G |a^{\dag}a|G \rangle  + \langle G |aa^{\dag}|G \rangle )^2 \nonumber \\&-4\langle G |(a^{\dag})^2|G \rangle\langle G |a^2|G \rangle\nonumber \\&= 4 \{ (\Delta x \Delta p)^2+\zeta\},
\end{align}
where we have notably used that $ \frac{e^{-E_0}}{1-e^{-\Delta}}=1$ since $\hbox{Tr}_{a} (\rho_r)= \langle G |G \rangle =1$ and where we have introduced $\zeta=(\langle G |a^2|G \rangle-\langle G |(a^{\dag})^2|G \rangle)^2/4$.
Finally, the pseudo-energy reads :
\begin{align}
\Delta=\log\left\{\frac{2\sqrt{(\Delta x \Delta p)^2+\zeta}+1}{2\sqrt{ (\Delta x \Delta p)^2+\zeta}-1}\right\}.
\end{align}

The derivation of the entanglement entropy becomes now straightforward:
\begin{align}
S&=-\hbox{Tr}_a\{\rho_r \log_2 (\rho_r)\}=-\hbox{Tr}_P\{\rho_r \log_2 (\rho_r)\}\nonumber\\
&=-\sum_{k=0}^{\infty} \langle k |(e^{-E_0-\Delta P^{\dag}P})\log_2(e^{-E_0-\Delta P^{\dag}P} ) |k\rangle\nonumber\\
&= \sum_{k=0}^{\infty} e^{-E_0-k\Delta}\frac{E_0+k\Delta}{\log{(2)}} \\
&=(\sqrt{(\Delta x\Delta p)^2+\zeta }\,+\,\frac{1}{2} ) \log_2\Big{\{}\sqrt{(\Delta x\Delta p)^2+\zeta}\,+\,\frac{1}{2}  \Big{\}}\nonumber\\& - (\sqrt{(\Delta x\Delta p)^2+\zeta}\,-\,\frac{1}{2})  \log_2\Big{\{}\sqrt{(\Delta x\Delta p)^2+\zeta}\,-\,\frac{1}{2}\Big{\}} \nonumber 
\end{align}  
 where the states $|k\rangle$ are Fock states for the $P$ operator.

{\it A priori}, the parameter $\zeta$ is  not always equal to 0 \cite{zeta}. 
 But on the other hand, it is always possible to come down to the case $\zeta = 0$ .
 In fact, if one starts from a quadratic Hamiltonian $\mathcal{H}(a,b_2,...,b_{n})$, with $\langle G |a^2|G \rangle = e^{2i\phi} | \langle G |a^2|G \rangle | $, (and  $2\phi \neq 0 [\pi]$), by introducing  $\tilde{a}=a e^{-i\phi}$, the formally new Hamiltonian $\tilde{\mathcal{H}}(\tilde{a},b_2,...,b_{n})=\mathcal{H}(a,b_2,...,b_{n})$, has a ground state $|\tilde{G}\rangle=\mathcal{N} e^{(\tilde{\Phi}^{\dag})^T \tilde{\mathcal{A}}\tilde{\Phi}^{\dag}} |0\rangle \otimes |0\rangle..|0\rangle=|G\rangle$, where $((\tilde{\Phi})^{\dag})^T=((\tilde{a})^{\dag},b_2^{\dag},...,b_{n}^{\dag}$), and where $\tilde{\mathcal{A}}=D(e^{-i\phi})\mathcal{A}D(e^{-i\phi})$ with $D(e^{-i\phi})$ the $n\times n$ diagonal matrix defined by $D(1,1)=e^{-i\phi}$ and $D(j,j)=1$ $\forall \, 2\leq j \leq n$.
Then $\langle \tilde{G} | \tilde{a}^2 |\tilde{G}\rangle = |\langle G| a^2 |G\rangle|$ and $4(\Delta\tilde{x})^2(\Delta\tilde{p})^2=-\langle G|[\tilde{a}+(\tilde{a})^{\dag}]^2|G\rangle\langle G|[\tilde{a}-(\tilde{a})^{\dag}]^2|G\rangle=4(\Delta x)^2(\Delta p)^2+(\langle G |a^2|G \rangle-\langle G |(a^{\dag})^2|G \rangle)^2$.
Finally, even if the introduction of a new photonic operator $\tilde{a}$ is needed, one can always convey to Eq. (\ref{relation}).
Note that for the Hamiltonians studied in the present article, i.e. the standard Dicke Hamiltonian (See Eq. (\ref{dickehamiltonian}) below) and the double quadrature Hamiltonian (See Eq. (\ref{hamiltoniandbl}) below)), one directly has $\zeta=0$, with no need to introduce $\tilde{a}$. 
In the case where $\Delta x\Delta p $ diverges, the Taylor expansion $ \log_2\{ \Delta x\Delta p \pm 1/2\}= \log_2\{ \Delta x\Delta p \}\pm [2\log(2)\Delta x\Delta p]^{-1}+{\it o}[(\Delta x\Delta p)^{-1}]$  implies Eq. (\ref{relentang}) : $S \sim \log_2\{ \Delta x\Delta p \}$, which provides a very simple relation between entanglement and fluctuations around the QCP.
Equivalently, the $\alpha$-R\'{e}nyi entropies,  defined as:
\begin{align}
S_{\alpha} (\rho_r)= \frac{1}{1-\alpha} \log_2\{\hbox{Tr}(\rho_r^{\alpha})\}
\end{align}
might also be written as:
\begin{align}
S_{\alpha} (\rho_r)= \frac{\alpha -\log_2\{(1+2\Delta x\Delta p)^{\alpha}-(2\Delta x\Delta p-1)^{\alpha}\}}{1-\alpha}.
\end{align}
  Note that the entanglement entropy corresponds to $S_{\alpha \rightarrow 1}$.
In the case where $\Delta x\Delta p $ diverges, the $\alpha$-R\'{e}nyi entropies also diverge as a logarithm : $ S_{\alpha} (\rho_r)\sim \log_2\{ \Delta x\Delta p \}.$

Below, to illustrate this relation between entanglement properties and the Heisenberg principle, we focus on standard models of photons and compute directly
the HP. 

 \section{The Dicke Model}
 \label{dicke}
We first focus on the Dicke Hamiltonian (DH), which describes the coupling of a single and uniform bosonic mode $a$ of energy $ \omega$ with $N$ two-level systems with atomic splitting $\omega_0$ (again, the Planck constant $\hbar$ has been fixed to unity for simplicity) \cite{Dicke}: 
\begin{eqnarray}
\label{dickehamiltonian}
H = \,\omega a^{\dag}  a \,+ \,\omega_0 J_z\,\,+ \frac{\lambda}{\sqrt{N}}(a + a^{\dag})(J_+\,+J_-),
\end{eqnarray} 
where $\lambda$ is the atom-field coupling strength.
The total angular momentum operators $J_z$ and $J_{\pm}$   read  $J_z=\sum_{l=1}^{N} \sigma_z^{l} $ and $J_{\pm}=\sum_{l=1}^{N} \sigma_{\pm}^{l}$, where $\sigma_z^{l}$ and $\sigma_{\pm}^{l}$ are the usual Pauli matrices for the ${l}^{th}$ pseudo-spin, so that the angular commutation relations are $[J_z,J_{\pm}]=\pm  2 J_{\pm}$ and $[J_+,J_-]=2 J_z$. In the thermodynamic limit ($N\rightarrow\infty$), this Hamiltonian undergoes a superradiant QPT for 
$\lambda=\lambda^{cr}=\sqrt{\omega\omega_0}/2$ \cite{Brandes}. When $\lambda<\lambda^{cr}$, the system is in a Normal Phase (NP), with a squeezed and non-degenerate vacuum, while the Superradiant Phase (SP) occurring for $\lambda>\lambda^{cr}$ is embodied by the appearance of a double degeneracy with atomic and photonic macroscopic coherences. This QPT is associated with the breaking of the parity operator $\Pi$ which accounts for the parity of the total number of excitation quanta: 
$\Pi=\exp(i\pi N_{exc})$ with $N_{exc}=a^{\dag}a+J_z+N/2$. 
 Moreover, through the Holstein-Primakoff transformation \cite{holstein}, which allows to write the angular operator in terms of a bosonic field $b$: $J_+ = b^{\dag} (N - b^{\dag} b)^{1/2}$ , $J_- = (N - b^{\dag}b)^{1/2}b$ and $J_z =b^{\dag} b\,-\,N/2$, the DH 
 can be proved \cite{Brandes} to be equivalent to  $H_{N\rightarrow\infty}= \tilde{\Delta}_{+}e_+^{\dag}e_++\tilde{\Delta}_{-}e_-^{\dag}e_-+E_G.$
 \begin{figure}

\begin{center}
\includegraphics[width=240pt]{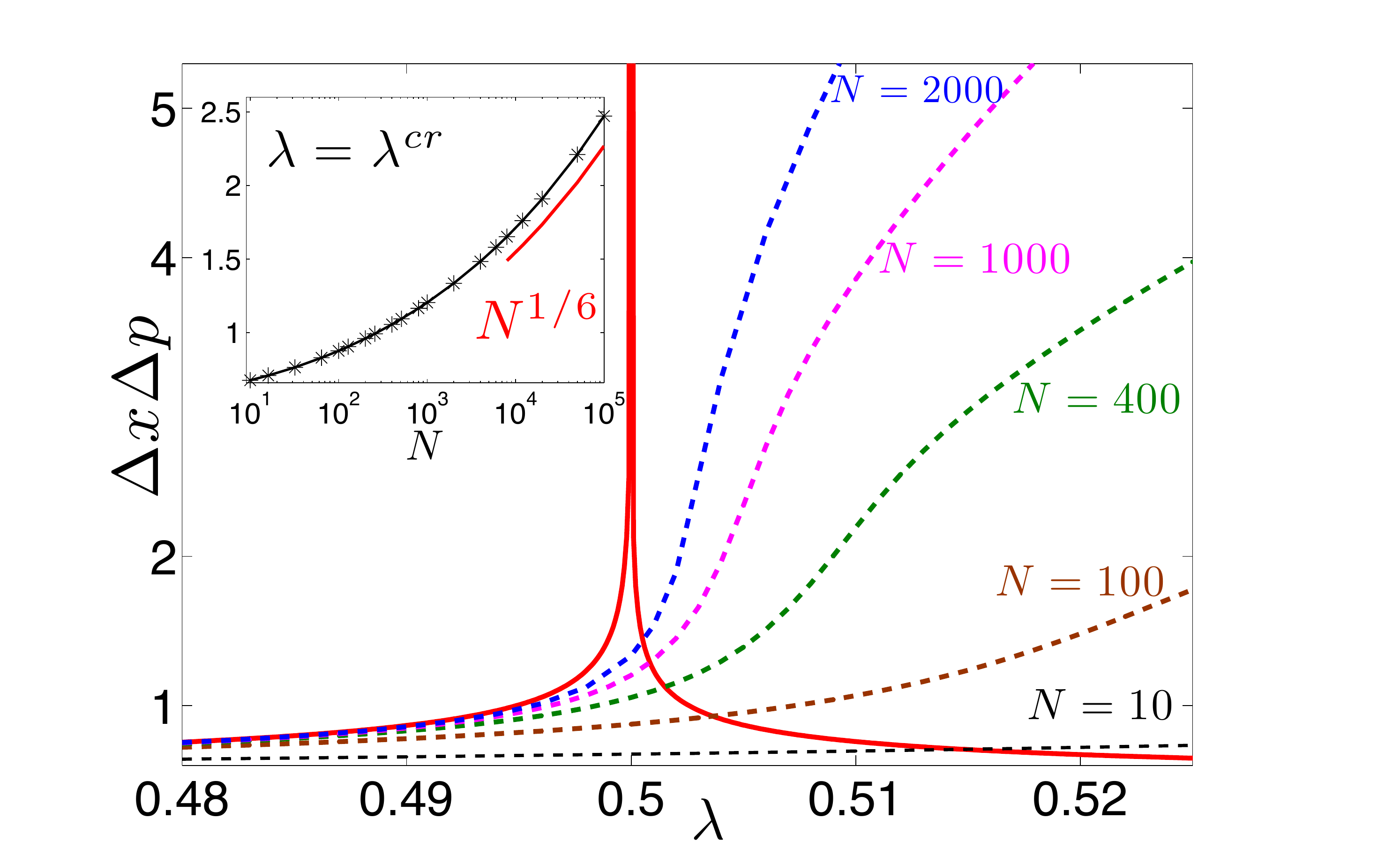}
\vskip -0.2cm
\caption{(Color online) In red, $\Delta x \Delta p$ (HP) for the ground state(s) of the Dicke model (cf Eq. (\ref{dickehamiltonian})) in the thermodynamic limit ($N\rightarrow \infty)$, with the Planck constant $\hbar$ fixed to unity. As a reminiscence of the entanglement entropy between atoms and light \cite{lambert}, it diverges at $\lambda=\lambda^{cr}$ as $|\lambda-\lambda^{cr}|^{-1/4}$. In dashed line, results given for the finite-size ground states of some exact diagonalizations. Since those are Schr\" odinger's cat like (restoring the broken symmetry), their HP diverge when $\lambda \gg\lambda^{cr}$, in contrast to the thermodynamic case. Inset: $\Delta x \Delta p$ at $\lambda=\lambda^{cr}$ versus  $N$ (the number of two-level systems), for $N=10$ to $N=10^5$; it scales like $N^{1/6}$ \cite{vidal,note}. Note that
the limit of $N=2.10^5$ $^{87}$Rb atoms coupled to an optical cavity has been achieved in BEC realizations of the Dicke model \cite{Zurich}. \label{fluctuations}}
\vspace{-0.5cm}
\end{center}
\end{figure}

 $E_G$ is the fundamental energy, and the normal eigenfrequencies (gaps) are such that  $2\tilde{\Delta}_{\pm}^2=(\omega_0/\mu)^2+\omega^2\pm\sqrt{((\omega_0/\mu)^2-\omega^2)^2+16\lambda^2\omega\omega_0\mu}$, with 
  $\mu=1$ if $\lambda<\lambda^{cr}$ and   $\mu=\frac{\omega_0\omega}{4\lambda^2}$ if $\lambda>\lambda^{cr}$. 
 The bosonic operators (the polaritons) $e_{\pm}$ are linear combinations of the original fields:
 \begin{eqnarray}
&e_-=\frac{1}{2}\Big{\{}\frac{\cos(\gamma)}{\sqrt{\omega\tilde{\Delta}_-}}[(\tilde{\Delta}_--\omega)(a^{\dag}+\alpha)+(\tilde{\Delta}_-+\omega)(a+\alpha)] \nonumber \\
&-\frac{\sin(\gamma)}{\sqrt{\tilde{\omega}_0\tilde{\Delta}_-}}[(\tilde{\Delta}_--\tilde{\omega}_0)(b^{\dag}-\beta)+(\tilde{\Delta}_-+\tilde{\omega}_0)(b-\beta)]\Big{\}} \,\,\,\,\,\, \label{lowerstandard}  \\
&e_+=\frac{1}{2}\Big{\{}\frac{\sin(\gamma)}{\sqrt{\omega\tilde{\Delta}_+}}[(\tilde{\Delta}_+-\omega)(a^{\dag}+\alpha)+(\tilde{\Delta}_++\omega)(a+\alpha)] \nonumber \\ 
&+\frac{\cos(\gamma)}{\sqrt{\tilde{\omega}_0\tilde{\Delta}_+}} [(\tilde{\Delta}_+-\tilde{\omega}_0)(b^{\dag}-\beta)+(\tilde{\Delta}_++\tilde{\omega}_0)(b-\beta)]\Big{\}},\,\,\, \,\,\,\label{upperstandard}
 \end{eqnarray}
where $\tilde{\omega}_0=\omega_0(\mu+1)/(2\mu)$, $\alpha=\epsilon  \sqrt{N(1-\mu^2)}\lambda/\omega$ is the photonic coherence, $\beta=\epsilon  \sqrt{N(1-\mu)/2}$ the electronic coherence, with $\epsilon=0$ in the NP and $\epsilon=\pm1$ in the SP. 
The mixing angle $\gamma$ is such that $\tan(2\gamma)=(4\lambda\sqrt{\omega\omega_0}\mu^{5/2})/(\omega_0^2-\mu^2\omega^2)$. The coherences are zero in the NP while they can  be either positive or negative in the SP, resulting in the double degeneracy of the eigenspectrum. In both cases, the ground state(s) $|G\rangle_{\epsilon}$, defined for a given set of coherences are the ones of a double harmonic oscillator (shifted in the SP), and satisfy 
$e_+|G\rangle_{\epsilon}=e_-|G\rangle_{\epsilon}=0$ \cite{detail}. Besides, since $\Pi=\exp(i\pi \{a^{\dag}a+b^{\dag}b\})$, then $\Pi a \Pi^{\dag}=-a$ and  $\Pi b \Pi^{\dag}=-b$. Consequently, in the SP,  $\Pi |G\rangle_{\pm}= |G\rangle_{\mp}$ , i.e. the symmetry $\Pi$ is broken : in contrast to the NP, the ground states $|G\rangle_{\pm}$ are no longer eigenstates of the parity operator $\Pi$. After inverting the last polaritonic relations,  one gets, for each ground state 
\begin{eqnarray}
&\Delta x \Delta p= \sqrt{\{ \frac{\cos(\gamma)^2}{2\tilde{\Delta}_-}+\frac{\sin(\gamma)^2}{2\tilde{\Delta}_+} \}\{  \frac{\cos(\gamma)^2\tilde{\Delta}_-}{2}+\frac{\sin(\gamma)^2\tilde{\Delta}_+}{2} \}},\,\,\,\,\,\,\,
\end{eqnarray} 

which is plotted in Fig. \ref{fluctuations} (in red).
For $\lambda \rightarrow \lambda^{cr}$,  $\tilde{\Delta}_1 \sim |\lambda-\lambda^{cr}|^{1/2}$  \cite{Brandes} and the HP diverges like  $|\lambda-\lambda^{cr}|^{-1/4}$. Thus, in the thermodynamic limit, where the DH in Eq. (\ref{dickehamiltonian})  is quadratic (see   $H_{N\rightarrow\infty}$ above),  the criticality of the ground state entanglement \cite{lambert} at the QCP is indeed involved by the divergence of the HP, as proved by Eq. (\ref{relentang}). 
 It is also important to compare the criticality given  by the mean-field with some finite size results.
First, in the  limit $\lambda/\omega \rightarrow \infty$ (the so-called  ultrastrong coupling limit \cite{ciuti,devoret,solano}), an $N^{th}$ order perturbative theory allows us to prove that the two first eigenstates $\vert \Psi_G \rangle$ and $\vert \Psi_E\rangle$, have their
energies  separated by an exponentially small splitting, and are linear superpositions of the states  $|\alpha_F \rangle\,  \vert N/2 \rangle_x$ and $|-\alpha_F \rangle\,  \vert -N/2 \rangle_x$.
Here, $| \pm\alpha_F \rangle$ are coherent states for the photonic part:
$a| \pm\alpha_F\rangle=\pm\alpha_F|\pm\alpha_F\rangle$ with $\pm\alpha_F=\pm\sqrt{N}\lambda/\omega$  \cite{degvacua,degvacuabis}. The states $| \pm N/2 \rangle_x$ are the two maximally polarized Dicke states in the x-direction (the direction of the coupling) : they  satisfy 
$J_x |\pm N/2\rangle =(1/2)(J_++J_-) |\pm N/2\rangle=\pm N/2|\pm N/2\rangle =\pm N/2 \Pi_{j=1}^{N} \vert \pm \rangle_j $ where each local pseudo-spin state $\vert \pm \rangle_j $ satisfies  $\sigma_x^j | \pm \rangle_j = \pm | \pm \rangle_j$.
The light-matter coupling is so important that each pseudo-spin is polarized in the direction of the coupling ( the $x$-direction in this paper).
Moreover, since $\Pi \vert \alpha_F \rangle\,  \vert N/2 \rangle_x=e^{i\pi a^{\dag}a}\vert \alpha_F \rangle \otimes \Pi^N_{j=1} e^{i\pi (\sigma_z^{j}+1/2)} |+\rangle_j  = (-1)^N\vert -\alpha_F \rangle\,  \vert- N/2 \rangle_x$, the following cat's states wave-functions \cite{degvacua2} 
are the only ones that restore the broken symmetry $\Pi$ :
\begin{eqnarray}
\label{finite}
\vert \Psi_G \rangle &\simeq \frac{1}{\sqrt{2}} \big{\{}\vert \alpha_F \rangle\,  \vert N/2 \rangle_x  +(-1)^N
\vert-\alpha_F \rangle\, \vert - N/2\rangle_x 
\big{\}} \nonumber\,\,\,\\
\vert \Psi_E\rangle &\simeq \frac{1}{\sqrt{2}} \big{\{} \vert \alpha_F \rangle\, \vert N/2 \rangle_x  -(-1)^N
\vert-\alpha_F \rangle\,  \vert -N/2 \rangle_x 
\big{\}}.\,\,\,
\end{eqnarray}

Then, from the expression of $\vert \Psi_G \rangle$ in Eq. (\ref{finite}) one identifies $\Delta x \Delta p \simeq(\alpha_F^2+1/4)^{1/2} \sim \sqrt{N}\lambda/\omega$ which  asymptotically matches the finite size curves of Fig. \ref{fluctuations} for $\lambda/\omega \gg 1$.
In fact, $\vert \Psi_G \rangle$ is a symmetric superposition of the two thermodynamical vacua $|G\rangle_{\pm}$ which are
the ground states of a double harmonic oscillator shifted around some macroscopic coherences. Since those coherences get infinitely far from each other when increasing both the number of atoms $N$ and the coupling $\lambda$, the fluctuations of such a superposition diverge also in this limit, contrary to the thermodynamic limit result. Finally, we must evaluate the fluctuations of the finite-size ground states at $\lambda=\lambda^{cr}$ to see whether it diverges, or not.
The {\it scaling hypothesis} which relates the exponent of $N$ and the power of $|\lambda-\lambda^{cr}|$ in the finite-size developments of every physical observables at the QCP  \cite{vidal}, allows us to  prove that :
\begin{eqnarray}
\Delta x \Delta p\sim N^{1/6} \,\,\,\,\,\hbox{for} \,\,\,\,\,\lambda=\lambda^{cr},
\end{eqnarray}
 in quantitative agreement with the numerical simulations, confirming the divergence of the HP when $N\rightarrow \infty$.

Actually, in the standard DH, the critical scaling of the HP comes from the criticality of $\Delta x$
because the quadrature $x$ is the one that interacts with the two-level systems. But what happens if the two quadratures $x=(1/(\sqrt{2\omega})) (a+a^{\dag})$
and $p=(i\sqrt{\omega/2}) (a-a^{\dag})$ are coupled to two different chains of atoms? 
\section{The Double Quadrature Dicke Model}
 \label{dbldicke}
Let us consider the double quadrature Dicke Hamiltonian which has recently been introduced \cite{dblDicke}:
\begin{align}
\label{hamiltoniandbl}
H &= \,\omega_{cav} a^{\dag}  a \,+ \,\omega^0_C J_z^C\,\,+ \,\omega^0_I J_z^I \,\,\,\,\\\nonumber &+ \frac{2\lambda_C}{\sqrt{N_C}}(a + a^{\dag})J_x^C\,+ i\frac{2\lambda_I}{\sqrt{N_I}}(a - a^{\dag}) J_x^I , \nonumber
\end{align}
where the  chain of two level systems  labeled by $C$ (resp. $I$), has an atomic transition frequency $\omega^0_C$ (resp. $\omega^0_I$ ) and is coupled to the quadrature $a+a^{\dag}$ (resp. $i[a-a^{\dag}]$) via the coupling strength constant $\lambda_C$ (resp. $\lambda_I$).

\begin{figure}
\begin{center}
\includegraphics[width=250pt]{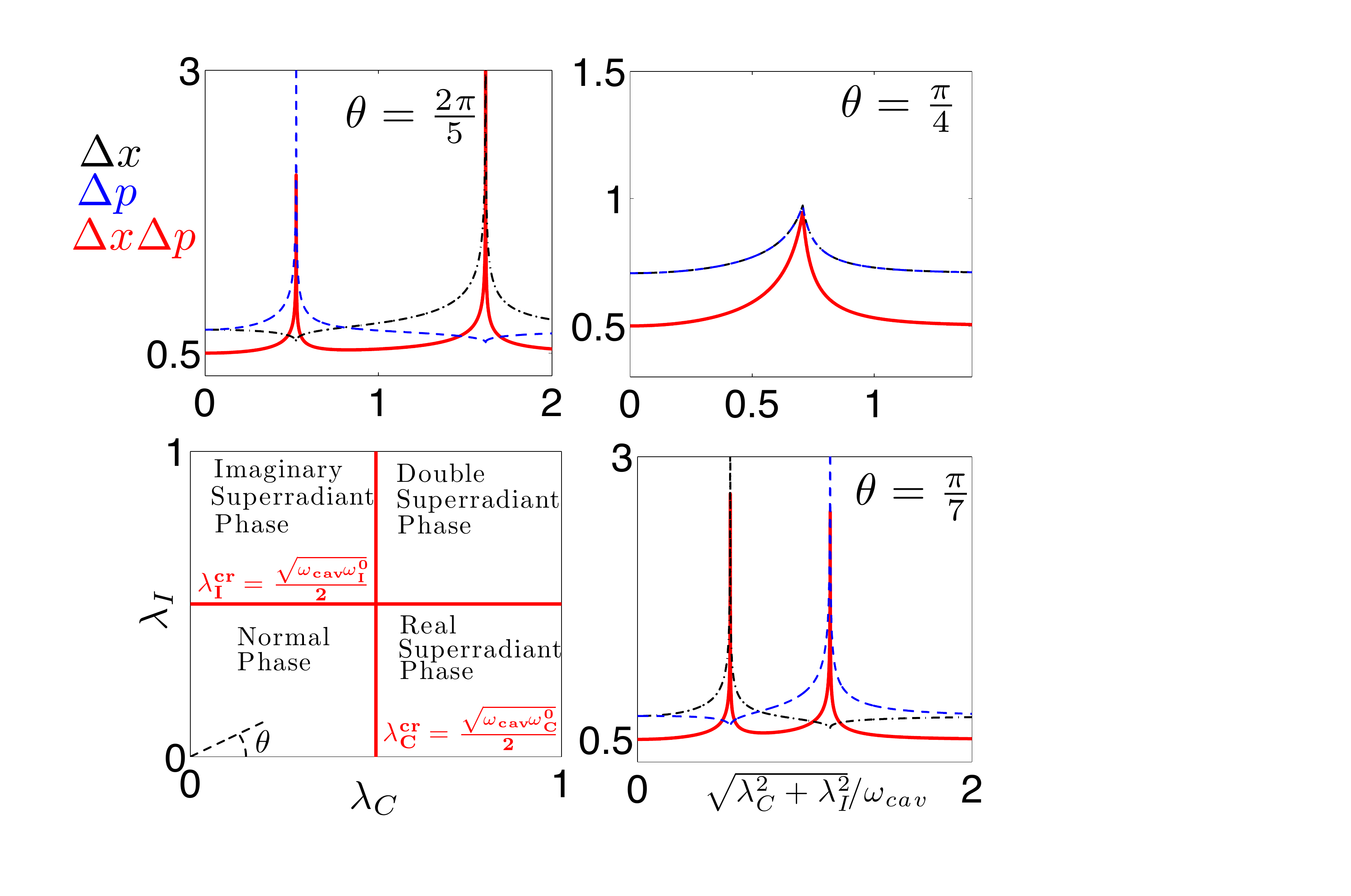}
\vspace{-0.4cm}
\caption{(Color online) Bottom left panel: two-dimensional phase diagram of the double quadrature Dicke model [see Hamiltonian (\ref{hamiltoniandbl})]. Other panels: fluctuations $\Delta x$ (black dashed dotted line),  $\Delta p$ (blue dashed line) and Heisenberg Principle $\Delta x\Delta p$ (solid red) in the resonant case ($\omega_{cav}=\omega_C^0=\omega_I^0=1$), in the thermodynamic limit ($N_C,N_I\rightarrow \infty $), and with the Planck constant $\hbar$ fixed to unity. The results are plotted with respect to the radial coupling $\sqrt{\lambda_C^2+\lambda_I^2}$,  with $\lambda_I/\lambda_C=cst=\tan(\theta)$,  for several polar angles $\theta$. At the QCPs, the product $\Delta x\Delta p$ shows either a divergence scaling as $1/\sqrt{\tilde{\Delta}_1}$, (where $\tilde{\Delta}_1$ is the lower gap) which is reminiscent of the one in the standard Dicke model, or a local maximum (at the point of double symmetry breaking, for $\theta=\pi/4$).  \label{fluctuationsdbl}}
\end{center}
\vspace{-0.4cm}
\end{figure}

Here, two independent symmetries transformations $\mathcal{T}_I$ and  $\mathcal{T}_C$ are conserved and are defined via the following operations:
\begin{eqnarray}
(a+a^{\dag},i(a-a^{\dag}),J_x^C,J_x^I)\stackrel{\mathcal{T}_I}{\rightarrow} (a+a^{\dag},-i(a-a^{\dag}),J_x^C,-J_x^I), \nonumber \,\,\,\,\,\\
(a+a^{\dag},i(a-a^{\dag}),J_x^C,J_x^I)\stackrel{\mathcal{T}_C}{\rightarrow} (-a-a^{\dag},i(a-a^{\dag}),-J_x^C, J_x^I)  \,\,\,\,\,\,\,\, \nonumber
\end{eqnarray}
where $ a^{\dag} a$, $J_z^I$ and $J_z^C$ remain unchanged. $\mathcal{T}_I$ can be viewed as
the time reversal symmetry \cite{dblDicke}. Again, by an Holstein-Primakoff transformation, one introduces the bosonic fields $b_C$ and $b_I$ defined via the relations: $J^k_+ = b^{\dag}_k (N_k - b^{\dag}_k  b_k)^{1/2}$ , $J^k_- = (N_k - b^{\dag}_k b_k)^{1/2} b_k$ and $J^k_z = b^{\dag}_k b_k\,-\,N_k/2$ ($k \in\{I,C\}$).  One can then show that $\mathcal{T}_C$ and  $\mathcal{T}_I$ gets broken when $\lambda_C$ and $\lambda_I$ are increased above $\lambda_C^{cr}=\sqrt{\omega_{cav}\omega^0_C}/2$ and $\lambda_I^{cr}=\sqrt{\omega_{cav}\omega^0_I}/2$ \cite{dblDicke}. This gives rise to four different quantum phases in the thermodynamic limit ($N_C,N_I\rightarrow \infty)$, separated by two orthogonal critical lines of equation $\lambda_C=\lambda_C^{cr}$ and $\lambda_I=\lambda_I^{cr}$, as shown in the bottom left panel of Fig. 
\ref{fluctuationsdbl}. One has either one (normal phase), two (real/imaginary superradiant phase) or four (double superradiant phase) degenerate and coherent vacua whose eigenfunctions are the ones of a triple harmonic oscillator :
 \begin{align}
 H_{N\rightarrow\infty}= \tilde{\Delta}_{1}e_1^{\dag}e_1+\tilde{\Delta}_{2}e_2^{\dag}e_2+\tilde{\Delta}_{3}e_3^{\dag}e_3+E_G^D.
\end{align}
 The gaps $\tilde{\Delta}_1\leq\tilde{\Delta}_2\leq\tilde{\Delta}_3$, the polaritons $e_1$, $e_2$ and $e_3$, and the fundamental energy $E_G^D$ are obtained by diagonalizing the associated Bogoliubov matrix \cite{dblDicke}.
\begin{figure}
\begin{center}
\hspace{-0.4cm}
\includegraphics[width=256pt]{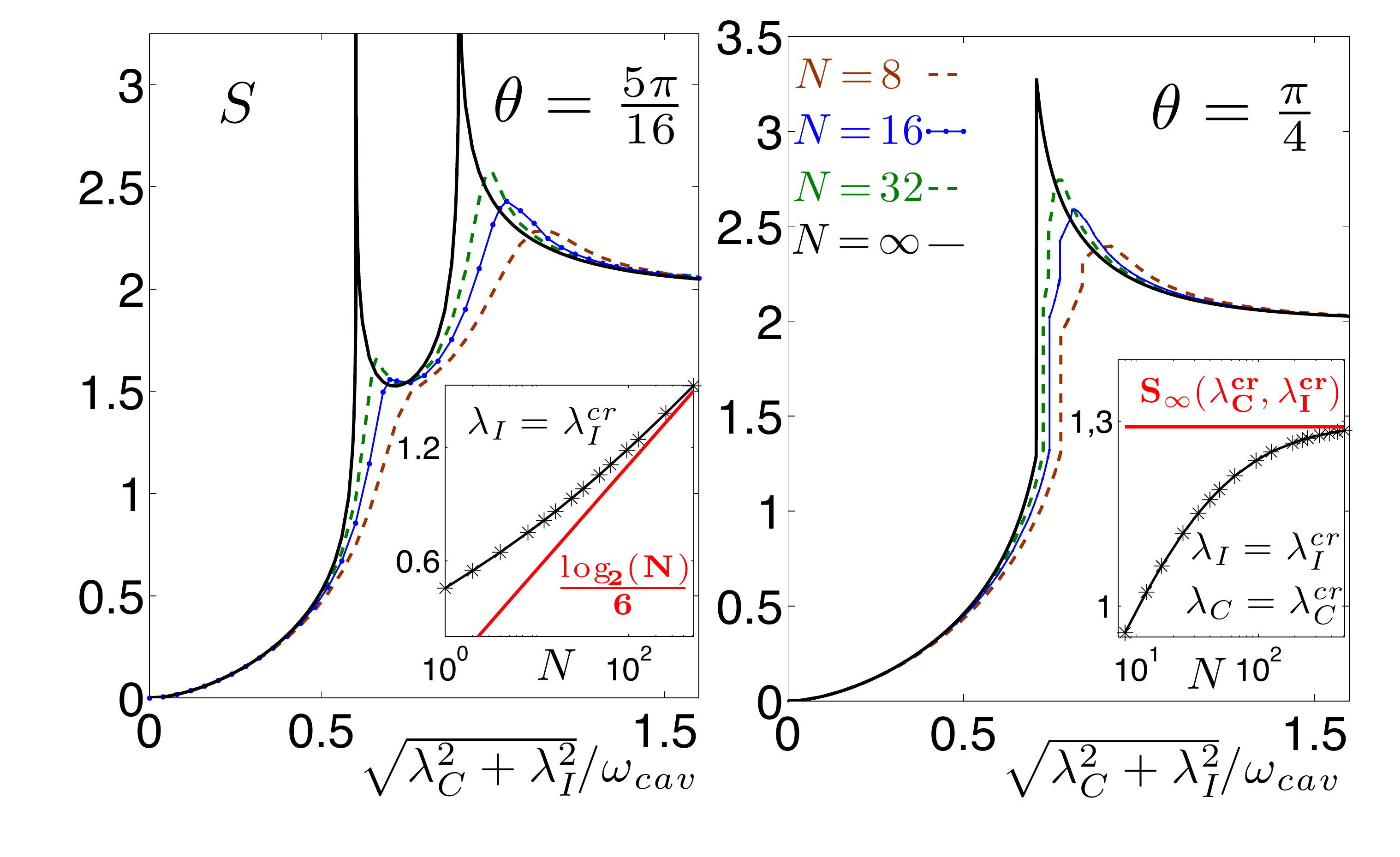}
\vspace{-0.4cm}
\caption{(Color online) Entanglement entropy  between the photonic field and the two matter fields 
$S=-\hbox{Tr}(\rho_r\log_2 \rho_r)$ (where $\rho_r=\hbox{Tr}_{b_C,b_I}\{|G\rangle \langle G|\}$) , with respect to $\sqrt{\lambda_C^2+\lambda_I^2}$, at resonance ($\omega_{cav}=\omega^0_C=\omega^0_I$) along the radial lines $\lambda_I/\lambda_C=\tan(\theta)$ for $\theta=5\pi/16$ (left),
and $\theta=\pi/4$ (right). Results shown for $N=N_C=N_I$, in the thermodynamic limit and in the finite size case. For $N=\infty$ and $\theta\neq\pi/4$, S diverges at the QCP like  $(-1/4)\log_2|\lambda_k-\lambda_k^{cr}|$ ($k=C,I$), i.e analogously to the standard Dicke model \cite{lambert}. For $\theta=\pi/4$ such a divergence disappears thanks to a compensation between the two simultaneous QPT (see Eq.(\ref{lowerdbl}) and (\ref{lowerpi4})). Insets: same quantity at the QCP, versus $N$ (up to N=512). For $\theta\neq\pi/4$, $S\sim\log_2(N)/6 $, as in the standard DH \cite{entangdickeN};
 for $\theta=\pi/4$, S does not diverge anymore: $S(N)\stackrel{N\rightarrow\infty}{\longrightarrow} S_{\infty}(\lambda_C^{cr},\lambda_I^{cr})\simeq 1.29$, providing
 a situation where a second order QPT admits a finite entanglement at its QCP.\label{entangdbl}}
\end{center}
\vspace{-0.6cm}
\end{figure}

We first study the case $\lambda_C \geq \lambda_I$, which corresponds to $\theta\leq\pi/4$ where $\tan(\theta)=\lambda_I/\lambda_C$ (see figure \ref{fluctuationsdbl}).
For $\lambda_C \rightarrow \lambda_C^{cr}$, and $\lambda_I < \lambda_I^{cr}$, the lower energy gap $\tilde{\Delta}_{1}$ vanishes as $|\lambda_C-\lambda_C^{cr}|^{1/2}$, and the lower polariton $e_1$ reads:
\begin{align}
\vspace{-0.3cm}
e_1 \simeq&\frac{1}{\mathcal{N}_{e_1}}
\Big{\{}(1+\frac{\omega_I^0\tilde{\Delta}_1}{\omega_I^0\omega_{cav}-4\lambda_I^2})\sqrt{\omega_C^0}\,a-(1+\frac{\tilde{\Delta}_1}{\omega_C^0})\sqrt{\omega_{cav}}\,b_C\,\,\,\,\,\nonumber\\&+2i\frac{\sqrt{\omega_C^0}\lambda_I\tilde{\Delta}_1}{\omega_I^0\omega_{cav}-4\lambda_I^2}b_I  
-(1-\frac{\omega_I^0\tilde{\Delta}_1}{\omega_I^0\omega_{cav}-4\lambda_I^2})\sqrt{\omega_C^0}\,a^{\dag}\nonumber\\&+(1-\frac{\tilde{\Delta}_1}{\omega_C^0})\sqrt{\omega_{cav}}\,b_C^{\dag}-2i\frac{\sqrt{\omega_C^0}\lambda_I\tilde{\Delta}_1}{\omega_I^0\omega_{cav}-4\lambda_I^2}b_I^{\dag}  \Big{\}} \label{lowerdbl},
\end{align}
where $\mathcal{N}_{e_1}=2\{\tilde{\Delta}_1(\frac{\omega_I^0\omega_C^0}{\omega_I^0\omega_{cav}-4\lambda_I^2}+\frac{\omega_{cav}}{\omega_C^0})\}^{1/2}$.

Those polaritonic coefficients exhibit the same divergence as the lower polariton $e_-$ of the standard DH (see Eq. (\ref{lowerstandard})).
In particular, the HP diverges for $\lambda_C \rightarrow \lambda_C^{cr}$ (and still $\lambda_I < \lambda_I^{cr}$) as $1/\sqrt{\tilde{\Delta}}_1=|\lambda_C-\lambda_C^{cr}|^{-1/4}$.
Interestingly, we realize by examining the expression (\ref{lowerdbl}),
that the polaritonic divergence disappears when  $\lambda_C \stackrel{{\small \lambda_C<\lambda_C^{cr}}}{\longrightarrow} \lambda_C^{cr}$ and $\lambda_I \stackrel{{\small \lambda_I<\lambda_I^{cr}}}{\longrightarrow} \lambda_I^{cr}$ simultaneously, which occurs only at the point of double symmetry breaking (at the crossing of the lines $\lambda_C=\lambda_C^{cr}$ and $\lambda_I=\lambda_I^{cr}$).
  At this point, corresponding to $\theta=\pi/4$ if $\omega_I^0=\omega_C^0$, the two matter modes play a symmetric 
  role and one has:
\begin{align}
&e_1 = a-\sqrt{\frac{\omega_{cav}}{4\omega_C^0}}\Big{(}b_C-i b_I\Big{)}+\sqrt{\frac{\omega_{cav}}{4\omega_C^0}}\Big{(}b_C^{\dag}-ib_I^{\dag}\Big{)} \label{lowerpi4}.
\end{align}

There, the fluctuations $\Delta x$ and $\Delta p$ do no longer diverge and 
$\Delta x \Delta p= 1/2+ (1+4(\omega_C^0/\omega_{cav})^2)^{-1/2}$.
As shown in Fig. \ref{fluctuationsdbl}, this value  is a local maximum of the HP along the line $\lambda_C=\lambda_I$. 

More generally, at every other QCPs of the two-dimensional phase diagram, 
where $\mathcal{T}_i$ or $\mathcal{T}_C$ are individually broken, the fluctuations of the quadrature involved in the transition diverge
as $1/\sqrt{\tilde{\Delta}_1}$, while the other quadrature fluctuations remained bounded.
Thus, at those single QCPs, one has either $\Delta x \sim |\lambda_C-\lambda_C^{cr}|^{-1/4}$ and $\Delta p < \infty$ for $\lambda_C\rightarrow\lambda_C^{cr}$ or $\Delta p \sim |\lambda_I-\lambda_I^{cr}|^{-1/4}$ and $\Delta x < \infty$ for $\lambda_I\rightarrow\lambda_I^{cr}$. 
Those scalings imply the critical behavior of the HP, which is the appropriate measure to detect all the QPTs in this model.\\
Finally, we could illustrate our initial statement about the equivalence of the measures of the HP and  the entanglement entropy S, by showing the behavior of this latter in this double quadrature model (See Fig. \ref{entangdbl}).
In order to compare the entanglement S in the thermodynamic limit to some finite-size situation, for which the ground states are cat's states in the superradiant phases,  one must add to the expression Eq. (\ref{relation}), a term accounting for the degeneracy \cite{lambert}, and equal to 1 (resp. 2) in the real or imaginary (resp. double) superradiant phases. That is why S saturates at 2 when $\lambda_C,\lambda_I \gg \omega_{cav}$. \\As expected, for $\theta\neq \pi/4$, at the single QCPs ($\lambda_C^{cr}$ or $\lambda_I^{cr}$), the divergence of the  HP ( due to the divergence of the polaritonic coefficients, see Eq. (\ref{lowerdbl})), involves the following scaling of the entanglement entropy  in the thermodynamic limit:

\begin{align}
S\sim (-1/4)\log_2|\lambda_k-\lambda_k^{cr}|,
\end{align}
 with $k \in\{C,I\}$.
This is reminiscent of the standard Dicke model  \cite{lambert}.
Moreover,  in the finite-size situation,  the entanglement entropy scales as  $S\sim (1/6)\log_2(N)$ (where $N=N_C=N_I$) at the single QCPs,  as an other reminiscence of the standard Dicke model \cite{entangdickeN}.  
In the left panel of fig. \ref{entangdbl}, we  show the plot of the entanglement S  along $\theta=5\pi/16$, both in the thermodynamical and finite-size cases. We clearly observe the two consecutive critical enhancement of S at the two consecutive QCPs.\\
On the other hand, at the point of double symmetry breaking (corresponding to $\theta=\pi/4$ at resonance $\omega_{cav}=\omega_C^0=\omega_I^0$) , the polaritonic coefficients do not diverge anymore (see Eq. (\ref{lowerpi4})).
Consequently, the entanglement S stays bounded in the thermodynamic limit, and apart from the discontinuity equal to 2 immediately after the double critical point $(\lambda_C^{cr},\lambda_I^{cr})$ ( due to the appearance of the four fold degeneracy), its value, given by Eq. (\ref{relation}), reads at resonance : 
\begin{align}
S_\infty(\lambda_C^{cr},\lambda_I^{cr})&=\frac{\{1+\sqrt{5}\}}{\sqrt{5}}\log_2(1+\sqrt{5})-\log_2(\sqrt{5})\nonumber\\
&\simeq 1.29.
\end{align}

 Correspondingly, for $N<\infty$ at this point, the physical quantities admit a standard $1/N$ finite-size expansion, and $S(N)\stackrel{\infty}{\rightarrow} S_{\infty}(\lambda_C^{cr},\lambda_I^{cr})< \infty$. Thus, while it undergoes a second order QPT, the entanglement of the system does not diverge when the number of pseudo-spins tends to infinity, which is somehow unusual, but in perfect agreement with the behavior of the HP.

\section{Conclusion}
 \label{conclusion}
   To summarize, in this work we have exemplified the enhancement of fluctuations at superradiant QPTs, through the HP. By exhibiting a general relation valid for any quadratic bosonic Hamiltonian (see Eq. (\ref{relation})), we have shown that the HP is indeed connected to the logarithmic enhancement of the entanglement entropy, while being certainly easier to measure, since it does not require the full tomographic determination of the density matrix \cite{tomography}, but just the measurements of the variance of the two orthogonal field quadratures. By the way, we would like to point out that the two models presented above could be physically implemented either with atomic Bose-Einstein condensates in optical cavities \cite{Zurich, nagy} or in circuit QED \cite{dblDicke,degvacua,cpbdicke}. For the latter proposal, the coupling to the quadrature $(a+a^{\dag})$ is provided by the {\it capacitive} coupling of the quantized charge of a Josephson atom and the quantum voltage of a resonator \cite{cpbdicke}, while the coupling to the quadrature $i(a-a^{\dag})$ is made thanks to the {\it inductive} coupling which connects the resonator current to the flux of the qubit \cite{devoret}. 

We acknowledge fruitful and stimulating discussions with C. Ciuti and J. Vidal. This work was supported by NSF under the grant DMR-0803200 and also by the DOE grant via DE-FG02-08ER46541.


\begin{thebibliography}{}
\bibitem{zeta} For instance, if one starts from the Hamiltonian $H_{Dicke}^{\eta} = \,\omega a^{\dag}  a \,+ \,\omega_0 J_z\,\,+ \frac{\lambda}{\sqrt{N}}(a e^{i\eta} + a^{\dag}e^{-i\eta})(J_+\,+J_-)$, with  $J_z$ and $J_{\pm}$ defined below Eq. (\ref{dickehamiltonian}) and $\eta \neq 0 [\pi]$, then $Im\{\langle G |a^2|G \rangle\}\neq 0$ and $\zeta\neq 0$.
\bibitem{detail}We should also add some indices $\epsilon$  to the operators $e_{\pm}$(and their {\bf h.c}), since those also depend on the macroscopic coherences (whose signs are given by $\epsilon$), but we choose to drop those for notational convenience.
\end{thebibliography}
  \end{document}